\documentclass[a4paper,12pt]{article}
\usepackage{epsfig}
\def\1{\'{\i}}
\begin{document}

\title{\bf The Limit Cycles of Li\'enard Equations
in the Weakly Nonlinear Regime}

\author{J.L. L\'{o}pez$^{\dag}$ and R. L\'{o}pez-Ruiz$^{\ddag}$ \\
                                  \\
{$^{\dag}$\small Department of Mathematics and Informatics,} \\
{\small Universidad P\'ublica de Navarra, 31006-Pamplona (Euroland).} \\
{$^{\ddag}$\small Department of Computer Science and BIFI,} \\
{\small Universidad de Zaragoza, 50009-Zaragoza (Euroland).}
\date{ }}

\small
\maketitle
\baselineskip 8mm
\begin{center} {\bf Abstract} \end{center}
Li\'enard equations of the form $\ddot{x}+\epsilon f(x)\dot{x}+x=0$, 
with $f(x)$ an even function, are considered in the weakly nonlinear regime 
($\epsilon\rightarrow 0$). A perturbative algorithm for obtaining 
the number, amplitude and shape of the limit cycles of these systems is given. 
The validity of this algorithm is shown and
several examples illustrating its application are given. 
In particular, an ${\mathcal O}(\epsilon^8)$ approximation for the amplitude 
of the van der Pol limit cycle is explicitly obtained.

$\;$\newline

\newpage

\section{Introduction}

A pattern formed by a continuum of periodic orbits is the usual
look of phase space for a conservative planar system. Take, for instance,
the harmonic oscillator $\ddot{x}+x=0$. The set of circles $x^2+\dot{x}^2=a^2$, 
with $a$ a positive real number representing the amplitude of the oscillation, 
fills the whole plane $(x,\dot{x})$. Nevertheless, 
dissipation and amplification are always present in real dynamics, since the oscillating system is 
usually embedded in an interacting external medium. The effect of this external medium
is implemented in the evolution equations of the oscillator, with more or less accuracy,
through a nonlinear term $h(x,\dot{x})$ that depends on the position and velocity of the oscillator.
In general, even if this new term is only a slight perturbation governed by the
small real parameter $\epsilon$,
the coexistence of the infinitely many periodic motions of the original unperturbed system is 
destroyed and only a finite number of them survives in the new context given by
the equation $\ddot{x}+x+\epsilon h(x,\dot{x})=0$.
Dissipation and amplification are balanced on those orbits. 
These localized periodic motions that verify strict energetic balance 
conditions are called {\it self-sustained oscillations} or 
{\it limit cycles} \cite{andronov,ye}. 

Li\'enard equations,
\begin{equation}
 \ddot{x}+\epsilon f(x)\dot{x}+x=0,
 \label{eq2}
\end{equation}
with $h(x,\dot{x})=f(x)\dot{x}$ and $f(x)$ any real function, 
model the dynamics of specific planar systems where limit cycles 
can be found. When $f(x)$ is a polynomial
of degree $N=2n+1$ or $2n$, with $n$ a natural number, 
Lins, Melo and Pugh have conjectured (LMP-conjecture)
that the maximum number of limit cycles allowed is just $n$ \cite{lins}.
It is true if $N=2$, or $N=3$
or if $f(x)$ is even and $N=4$ \cite{lins,rychkov}.
Also, it has been claimed its truth 
in the strongly nonlinear regime
$(\epsilon\rightarrow\infty)$ when $f(x)$ is an even
polynomial \cite{lopez}. There are no general results
about the limit cycles when $f(x)$ is a polynomial
of degree greater than $5$ neither, in general, when $f(x)$
is an arbitrary real function \cite{giacomini}.
When $f(x)=x^2-1$ we have the van der Pol oscillator, 
$\ddot{x}+\epsilon (x^2-1)\dot{x}+x=0$, wich
is a particular example of Li\'enard system that has been very well studied.
It displays a limit cycle whose uniqueness and non-algebraicity
has been shown for the whole range of the parameter $\epsilon$ \cite{odani}.
Its behavior runs from near-harmonic oscillations when $\epsilon\rightarrow 0$ 
to relaxation oscillations when $\epsilon\rightarrow\infty$, making
it a good model for many practical situations
\cite{lopezruiz}.

Different non-perturbative approaches that allow to obtain information about 
the number of limit cycles and their location in phase space have 
been proposed in the last years when $f(x)$ is an even polynomial.
A method that gives a sequence of algebraic approximations to the equation
of each limit cycle can be found in \cite{giacomini}, and a variational method
showing that limit cycles correspond to relative extrema of certain functionals
is explained in \cite{depassier}.

In this paper, we are interested in determining  
the number, amplitude and shape of the limit cycles emerging in the weakly 
nonlinear regime $(\epsilon\rightarrow 0)$ of Li\'enard equations
by standard perturbation techniques.
This work completes a previous work \cite{lopez} in which these questions
were successfully answered and solved in the strongly nonlinear 
regime $(\epsilon\rightarrow \infty)$.
In fact, the algorithm here explained is 
an alternative to a more general,
although technically more sophisticated, perturbative method 
proposed by Giacomini et al. \cite{giacomini1,giacomini2} on the same question.
Thus, we propose a new algorithm which permits an easy computation 
up to an arbitrary order ${\mathcal O}(\epsilon^N)$ of the amplitude $a$ and form $y(x)$ 
of the limit cycles. Moreover, this simple algorithm permits to show some interesting 
properties of symmetry of the approximate solutions. The article is organized as follows.
A few symmetries of the Li\'enard equation are recalled in Section 2. 
The method is detailed in Section 3.  Some illustrative examples are given in Section 4.
Finally, we present our conclusions.

\section{Symmetries and an integral equation}

In order to study the limit cycles of equation (\ref{eq2}) with the time
variable being implicit,
it is convenient to rewrite it in the coordinates $(x,\dot{x})=(x,y)$ in the plane, 
with $\dot{x}(t)=y(x)$ and $\ddot{x}(t)=y(x)y'(x)$ (where $y'(x)=dy/dx$):
\begin{equation}
 yy'+\epsilon f(x)y+x = 0.
 \label{eq3}
\end{equation}
A limit cycle $C_l\equiv (x,y_{\pm}(x))$ of equation (\ref{eq3})
has a positive branch $y_+(x)>0$ and a negative branch $y_-(x)<0$.
They cut the $x$-axis in two points $(-a_-,0)$ and $(a_+,0)$
with $a_-,a_+>0$ because
the origin $(0,0)$ is the only fixed point of Eq. (\ref{eq3}).
Then every limit cycle $C_l$ solution of Eq. (\ref{eq3}) encloses
the origin and the oscillation $x$ runs in the interval
$-a_-<x<a_+$. The amplitudes of oscillation $a_-,a_+$ identify
the limit cycle. The result is a nested set of closed curves
that defines the qualitative distribution
of the integral curves in the plane $(x,y)$. The stability
of the limit cycles is alternate. For a given stable limit cycle,
the two neighboring  limit cycles, the closest one in its interior
and the closest one in its exterior, are unstable,
and vice versa.

When $f(x)$ is an even function, the {\it inversion symmetry}
$(x,y)\leftrightarrow (-x,-y)$ of Eq. (\ref{eq3}) implies
$y_+(x)=-y_-(-x)$ and then $a_1=a_2=a$.
Therefore we can restrict ourselves to the positive branches
of the limit cycles $(x,y_+(x))$ with $-a\leq x \leq a$. In this case,
the amplitude $a$ identifies the limit cycle.
The {\it parameter inversion symmetry}
$(\epsilon,x,y)\leftrightarrow (-\epsilon,x,-y)$
implies that if $C_l\equiv (x,y_{\pm}(x))$ is a limit
cycle for a given $\epsilon$, then
$\overline{C}_l\equiv (x,-y_{\mp}(x))$ is a limit cycle
for $-\epsilon$. In consequence, the amplitude $a$ of the limit cycles
in Li\'enard systems is an even function of $\epsilon$.
Moreover if $C_l$ is stable (or unstable)
then $\overline{C}_l$ is unstable
(or stable, respectively). Therefore it is enough
to consider the limit
cycles when $\epsilon>0$ for obtaining all of the periodic solutions.
(The limit cycles for a given
$\epsilon <0$ are obtained from a reflection over
the $x$-axis of those limit cycles obtained for $\epsilon>0$).

Another property of a limit cycle can be derived from
the fact that the mechanical energy $E=(x^2+y^2)/2$ is
conserved in an oscillation:
\begin{displaymath}
\int_{C_l}\frac{dE}{dx} dx = 0.
\end{displaymath}
Taking into account that $dE=(yy'+x)\,dx=-\epsilon f(x)y\,dx$,
if equation (\ref{eq3}) is integrated along the positive
branch of a limit cycle, between the maximal amplitudes of oscillation,
we obtain:
\begin{equation}
\int_{-a}^{a}f(x)y_+(x) dx= 0.
\label{eqq3}
\end{equation}
The solutions $y_+(x)$ of Eqs. (\ref{eq3}) and (\ref{eqq3})
and vanishing in the extremes, constitute the finite set of limit 
cycles of Eq. (\ref{eq3}).

\section{The perturbative method}

For later convenience, we rewrite the Li\'enard equation (\ref{eq3}) with a 
different notation:
\begin{equation}
\label{lienard}
uu'+\epsilon f(t)u+t= 0,
\end{equation}
where $f(t)$ is an even function of the real variable $t$ and $u'$ means derivative 
with respect to $t$. From the discussion of the above section we know that the amplitudes 
of the limit cycles are symmetric: if $u(t)$ is a limit cycle of amplitude $a$ 
then $u(-a)=u(a)=0$. After the following change of dependent and independent variables:
\begin{equation}
\label{change}
t=ax, \hskip 2cm u=ay, \hskip 2cm a>0,
\end{equation}
equation (\ref{lienard}) reads
\begin{equation}
\label{liena}
yy'+\epsilon f(ax)y+x= 0,
\end{equation}
where $a$ is now an explicit positive parameter of the equation,
and $y'$ means now derivative with respect to $x$.
Every limit cycle $u(t)$ of (\ref{lienard}) of amplitude $a$ is 
transformed into a limit cycle solution $y(x)$ of (\ref{liena}) 
of amplitude $1$ verifying $y(-1)=y(1)=0$. The main result of this paper
is to establish in these new variables $(x,y)$ 
a recursive algorithm capable of approximating
a limit cycle solution $y(x)$ and its amplitude $a$ in a power series of 
$\epsilon$ up to an arbitrary order. The approximated limit cycles 
in the original variables $(u,t)$ are obtained by undoing the change of 
variables (\ref{change}) at each order ${\mathcal O}(\epsilon^N)$ 
of approximation. In order to show our results we need some previous
definitions.

\noindent                                                   
{\bf Definition 1.}
For $N=0,1,2,...$, we denote by $y^{(N)}(x)$ the function that approximates a limit cycle solution
$y(x)$ of (\ref{liena}) up to the order ${\mathcal O}(\epsilon^N)$:
\begin{equation}
\label{yN}
y^{(N)}(x)\equiv \sum_{n=0}^{N} \epsilon^ny_n(x).
\end{equation}
We denote by $a^{(N)}$ the approximation to its amplitude $a$ up to the order ${\mathcal O}(\epsilon^N)$:
\begin{equation}
\label{aN}
a^{(N)}\equiv \sum_{n=0}^{N-1} a_n\epsilon^n.
\end{equation}
The coefficient functions $y_n(x)$ and the coefficients $a_n$ are computed by means of the algorithm given below.

\noindent{\bf Definition 2.} For $n=0,1,2,...,N$, we denote by $f_n(a_0,a_1,...,a_n;x)$ 
the coefficients of the formal Taylor expansion of the function  
$g(\epsilon)\equiv f(a^{(N)}x)$ at $\epsilon=0$:
$$
f\left(x\sum_{n=0}^{N}a_n\epsilon^n\right)=\sum_{n=0}^N 
f_n(a_0,a_1,...,a_n;x)\epsilon^n+{\mathcal O}(\epsilon^{N+1}).
$$
The first few coefficients $f_n(a_0,a_1,...,a_n;x)$ are:
$$
f_0(a_0;x)=f(a_0x),
$$
$$
f_1(a_0,a_1;x)=x f'(a_0x)a_1,
$$
$$
f_2(a_0,a_1,a_2;x)=x f'(a_0x)a_2+{1\over 2}x^2 f''(a_0x)a_1^2,
$$
$$
f_3(a_0,a_1,a_2,a_3;x)=x f'(a_0x)a_3+x^2 f''(a_0x)a_1a_2+{1\over 6}x^3 f'''(a_0x)a_1^3.
$$

The general form of the coefficients $f_n(a_0,a_1,...,a_n;x)$ is given in 
the following lemma whose proof is straightforward and we do not reproduce here: 

\noindent                                                        
{\bf Lemma 1.}
For $n=0$ we have $f_0(a_0;x)= f(a_0x)$ and, for $n=1,2,...$, 
\begin{equation}
\label{efen}
f_n(a_0,a_1,...,a_n;x)=\sum_{m=1}^nb_{n,m}(a_1,...,a_n)x^mf^{(m)}(a_0x),
\end{equation}
where
$$
b_{n,m}(a_1,...,a_n)\equiv\sum_{S_{n,m}}{a_1^{\sigma(1)}a_2^{\sigma(2)}
\cdot\cdot\cdot a_n^{\sigma(n)}\over\sigma(1)!\sigma(2)!\cdot\cdot\cdot \sigma(n)!}.
$$
In these formulas $n\ge m\ge 1$ and the above sum is performed over the following subset $S_{n,m}$ of the 
set of all of the possible mappings $\sigma$ between the sets $\{1,2,3,...,n\}$ 
and $\{0,1,2,3,...,m\}$:
$$
S_{n,m}\equiv\left\{\sigma:\{1,2,3,...,n\}\to\{0,1,2,3,...,m\}; \hskip 2mm \sum_{k=1}^n\sigma(k)=m,
\hskip 2mm \sum_{k=1}^n k\sigma(k)=n\right\}.
$$
We also define $b_{0,0}=1$ and $b_{n,0}=0$ for $n>0$.

\noindent                                                        
{\bf Definition 3.}
Suppose that $y_0(x)$, $y_1(x)$,..., $y_n(x)$ are given. 
For $n=0,1,2,...$, we define the functions:
\begin{eqnarray}
\label{betan}
\beta_n(a_0,a_1,...,a_n) & \equiv & \sum_{m=0}^n\int_{-1}^1 
f_m(a_0,a_1,...,a_m;x)y_{n-m}(x)dx= \nonumber \\ 
& = & \sum_{m=0}^n\sum_{k=0}^mb_{m,k}\int_{-1}^1 x^ky_{n-m}(x)f^{(k)}(a_0x)dx.
\end{eqnarray}

\noindent{\bf Algorithm.} {\it We compute the coefficients $(y_n,a_n)$, $n=0,1,2,...,N$, 
of expansions (\ref{yN}) and (\ref{aN}) in the following way.

\noindent\underline{\bf Step 1}: Set $y_0(x)=\sqrt{1-x^2}$ and 
take one of the solutions $a_0$ of the (in general nonlinear) equation $\beta_0(a_0)=0$:
\begin{equation}
\beta_0(a_0)\equiv\int_{-1}^1 f_0(a_0x)y_0(x)dx=0.
\end{equation}

\noindent\underline{\bf Step 2}: For each $n$, with $n=1,2,...,N$, 
compute, firstly, $y_n(x)$  by means of the formula
\begin{eqnarray}
\label{yn}
y_n(x)\hskip -2mm &=&\hskip -2mm -{1\over y_0(x)} \left\lbrace{1\over 2}\sum_{m=1}^{n-1}y_m(x)y_{n-m}(x)+
\sum_{m=0}^{n-1}\int_{-1}^x y_{n-m-1}(t)f_m(a_0,a_1,...,a_m;t)dt\right\rbrace=  \nonumber \\ 
&=&\hskip -2mm -{1\over y_0(x)} \left\lbrace{1\over 2}\sum_{m=1}^{n-1}y_m(x)y_{n-m}(x)+
\sum_{m=0}^{n-1}\sum_{k=0}^mb_{m,k}\int_{-1}^x t^ky_{n-m-1}(t)f^{(k)}(a_0t)dt\right\rbrace
\end{eqnarray}
and, secondly, solve for $a_n$ the linear equation
\begin{equation}
\label{betann}
\beta_n(a_0,a_1,...,a_n)=0.
\end{equation}
}

The derivation of this algorithm is given in the following theorem. 
Its detailed form for $N=5$ is given in Section \ref{sec-ex}.

\noindent
{\bf Theorem 1.}
For $N=0,1,2,...$, the function $y^{(N)}(x)$ and the amplitude $a^{(N)}$
obtained by means of the above algorithm satisfy (\ref{liena}) 
up to the order ${\mathcal O}(\epsilon^{N+1})$ uniformly in $x\in[-1,1]$: 
\begin{equation}
\label{order}
y^{(N)}{dy^{(N)}\over dx}+\epsilon f(a^{(N)}x)y^{(N)}+x= {\mathcal O}(\epsilon^{N+1}).
\end{equation}
Moreover:

\noindent (i) All of the coefficients $y_n(x)$ of $y^{(N)}(x)$ are continuous 
functions of $x$ and satisfy $y_n(-1)=y_n(1)=0$ and therefore  
$y^{(N)}(-1)=y^{(N)}(1)=0$ $\forall$ $N$. 

\noindent (ii) If $\beta_0'(a_0)\ne 0$ 
(limit cycles of multiplicity higher than one are not considered here) 
then $a_{2n+1}=f_{2n+1}(a_0,a_1,...,a_{2n+1};x)=0$ for $n=0,1,2,...$. 
The coefficients $y_{2n+1}(x)$ are odd functions of $x$ and $y_{2n}(x)$ 
are even functions of $x$.  

\noindent (iii) When $f(x)$ is a polynomial of degree $q$ in $x^2$, then 
$$
y_{2n}(x)=p_{2n}(x^2)(1-x^2)^{3/2}, \hskip 5mm n=1,2,3,...,
$$ 
$$
y_{2n+1}(x)=xp_{2n+1}(x^2)(1-x^2), \hskip 5mm n=0,1,2,...,
$$
where $p_n(x)$ is a polynomial of degree $nq-1$. 

\noindent
{\bf Proof.}
If we replace $y$ by $y^{(N)}$ and $a$ by $a^{(N)}$ in (\ref{liena}) and 
equate powers of $\epsilon$ we obtain that
\begin{equation}
\label{yo}
y_0y_0'+x=0
\end{equation}
and, for $n=1,2,3,...$, every function $y_n(x)$ satisfies a first order linear differential 
equation which contains $y_0$, $y_1$,...,$y_{n-1}$:
\begin{equation}
\label{yene}
{1\over 2}\sum_{m=0}^n(y_my_{n-m})'+\sum_{m=0}^{n-1}f_m(a_0,a_1,...,a_m;x)y_{n-m-1}=0.
\end{equation}
The solution of (\ref{yo}) which satisfies $y_0(-1)=0$ is obviously $y_0(x)=\sqrt{1-x^2}$.
We solve recursively every one of the equations (\ref{yene}) for $y_n(x)$, $n=1,2,...,N$, 
in terms of the preceding functions $y_0$, $y_1$,..., $y_{n-1}$ and of the coefficients $a_0$, 
$a_1$,..., $a_{n-1}$ by imposing that $y_n(-1)=0$ for $n=1,2,3,...,N$. Then we obtain (\ref{yn}). 
Therefore, $y^{(N)}$ satisfies (\ref{liena}) with $a$ replaced by $a^{(N)}$ up to the order 
${\mathcal O}(\epsilon^{N+1})$ and (\ref{order}) holds. 

We already have that $y_n(-1)=0$ for $n=0,1,2,...,N$. To show thesis (i),
it remains to show that $y_n(1)=0$ for $n=0,1,2,...,N$
when Eq. (\ref{betann}) is assumed. We write (\ref{yn}) in the form
\begin{equation}
\label{ynb}
y_n(x)=w_n(x)-{\beta_{n-1}(a_0,a_1,...,a_{n-1})\over y_0(x)},
\end{equation}
with
\begin{equation}
\label{ynbb}
w_n(x)\equiv-{1\over y_0(x)} \left\lbrace{1\over 2}\sum_{m=1}^{n-1}y_m(x)y_{n-m}(x)+
\sum_{m=0}^{n-1}\sum_{k=0}^mb_{m,k}\int_x^1 t^ky_{n-m-1}(t)f^{(k)}(a_0t)dt\right\rbrace
\end{equation}
and $\beta_n(a_0,a_1,...,a_n)$ given in (\ref{betan}).
Obviously, $y_0(x)={\mathcal O}(\sqrt{1-x})$ when $x\to 1$ and is a 
continuous function in $[-1,1]$. 
From the hypothesis (\ref{betann}), we have that $\beta_{n}(a_0,a_1,...,a_{n})=0$ 
for $n=0,1,2,...,N-1$ and then $y_n(x)=w_n(x)$ 
for $n=1,2,3,...,N$. From (\ref{ynbb}), it is straightforward to show by induction over $n$ 
that all of the functions $w_n(x)$ are continuous in $[-1,1]$ and 
$w_n(x)={\mathcal O}(\sqrt{1-x})$ when $x\to 1$. Therefore, $y_n(x)$ is continuous 
in $[-1,1]$ and $y_n(1)=0$ for $n=0,1,2,...,N$. Then thesis (i) holds.

From Definition 2, all of the odd functions $f_{2n+1}(a_0,a_1,...,a_{2n+1};x)$ 
vanish if the odd coefficients $a_{2n+1}$ vanish. This is so because it is necessary 
to have a couple $(k,\sigma(k))$ with odd $k$ and $\sigma(k)$ in the definition of $b_{n,m}$ in
order to satisfy the condition $\sum_{k=1}^{2n+1} k\sigma(k)=2n+1$. 
Then, to show the first part of (ii) it suffices to show that $a_{2n+1}=0$, 
with $n=0,1,2,...$.

The function $y_0(x)$ is even and from (\ref{yn}) with $n=1$:
\begin{equation}
\label{yuno}
y_1(x)=-{1\over y_0(x)}\int_{-1}^x f(a_0 t)y_0(t)dt.
\end{equation}
Taking into account that $\beta_0(a_0)=\int_{-1}^1 f(a_0 t)y_0(t)dt=0$
and that $f(x)$ is even, we have $\int_{-1}^0 f(a_0 t)y_0(t)dt={1 \over 2}\beta_0(a_0)=0$.
We can write the integral in (\ref{yuno}) in the form $\int_{-1}^xf(a_0 t)y_0(t)dt= \int_{-1}^0f(a_0 t)y_0(t)dt+\int_{0}^xf(a_0 t)y_0(t)dt=\int_{0}^xf(a_0 t)y_0(t)dt$, which shows that $y_1(x)$ is odd. 
From (\ref{betan}) with $n=1$ we have
$$
\beta_1(a_0,a_1)=\int_{-1}^1 f(a_0 x)y_1(x)dx+a_1\int_{-1}^1 xf'(a_0 x)y_0(x)dx.
$$
The first integral vanishes because its integrand is odd and the second 
one equals $\beta_0'(a_0)$, which is not zero from assumption. 
Therefore, $\beta_1(a_0,a_1)=0\Rightarrow a_1=0$ and $f_1(a_0,a_1;x)=0$. 
Then, thesis (ii) is true for $n=0$. Suppose that it is true for $n=0,1,2,...,m-1$ with $m>0$; 
we will show that thesis (ii) is also true for $n=m$. If thesis (ii) is true for $n=0,1,2,...,m-1$, then
for $n\le m$, the integrals in the first line of (\ref{yn}) may be written in the form
\begin{eqnarray}
\label{suma}
&&\sum_{k=0}^{\lfloor (n-1)/2\rfloor}\int_{-1}^x y_{n-2k-1}(t)f_{2k}(a_0,a_1,...,a_{2k};t)dt=\nonumber\\
&&=\sum_{k=0}^{\lfloor (n-1)/2\rfloor}\int_{-1}^0 y_{n-{2k}-1}(t)f_{2k}(a_0,a_1,...,a_{2k};t)dt+\nonumber \\
&&\sum_{k=0}^{\lfloor (n-1)/2\rfloor}\int_0^x y_{n-{2k}-1}(t)f_{2k}(a_0,a_1,...,a_{2k};t)dt.
\end{eqnarray}
If $f(x)$ is an even function of $x$, then all of the functions $f_{2k}(a_0,a_1,...,a_{2k};x)$ are even. 
When $n$ is even then $y_{n-2k-1}(t)$, $k=0,1,2,...,\lfloor (n-1)/2\rfloor$, is odd and 
the function $\int_{-1}^x y_{n-2k-1}(t)f_{2k}(a_0,a_1,...,a_{2k};t)dt$ are even functions of $x$.
On the other hand, when $n$ is odd then $y_{n-2k-1}(t)$, $k=0,1,2,...,\lfloor (n-1)/2\rfloor$, 
are even and the first sum in the right hand side of (\ref{suma}) 
equals $\beta_{n-1}(a_0,a_1,...,a_{n-1})/2=0$ and then the left hand side is an odd function of $x$. 
Obviously, the sum $\sum_{k=1}^{n-1}y_ky_{n-k}$ in (\ref{yn}) is an even function of $x$ if $n\le m$ 
is even and an odd function of $x$ if $n\le m$ is odd. Then, the second part of thesis (ii) is true for $n=m$. 

From the induction hypothesis, $a_{2n+1}=f_{2n+1}=0$ for $n=0,1,2,...,\lfloor m/2\rfloor-1$ and then, 
for odd $n=m$, equation (\ref{betan}) reads:
\begin{equation}
\label{odbeta}
\beta_m(a_0,a_1,...,a_m)=\int_{-1}^1 f_m(a_0,a_1,...,a_m;x)y_0(x)dx.
\end{equation}
From Definition 2, for odd $m$, the function $f_m(a_0,a_1,...,a_m;x)$ is of the form
$f_m(a_0,a_1,...,a_m;x)=xf'(a_0x)a_m+g_m(a_1,a_3,...,a_{m-2};x)$, where 
$g_m(a_1,a_3,...,a_{m-2};x)=0$ when $a_1=a_3=...=a_{m-2}=0$. Then, for odd $n=m$ equation 
(\ref{odbeta}) may be written in the form
$$
\beta_m(a_0,a_1,...,a_m)=a_m\beta_0'(a_0).
$$
Therefore, the equation
$\beta_m(a_0,a_1,...,a_m)=0\Rightarrow a_m=0$ (and then $f_m=0$) for odd $m$. Then the first 
part of thesis (ii) is true for $n=m$.

To show (iii) we recall the function $y_1$ given in (\ref{yuno}).
If $f(x)$ is a polynomial of degree $q$ in $x^2$, then the integrand involved in the 
computation of $y_1(x)$ is a product of a polynomial of degree $q$ in $t^2$ by $\sqrt{1-t^2}$. 
An easy computation shows that
$$
\int_{-1}^x f(a_0 t)\sqrt{1-t^2}=\alpha+xp(x^2)\sqrt{1-x^2}+\beta\arcsin(x).
$$
where $\alpha$ and $\beta$ are real numbers and $p(x)$ a polynomial of degree $q$. 
But $y_1(\pm 1)=0$ and then
$$
\int_{-1}^1 f(a_0 t)\sqrt{1-t^2}=\alpha\pm\beta\arcsin(1)=0\Rightarrow \alpha=\beta=0.
$$
Moreover, $y_1(x)$ is odd and from (\ref{yuno}) and $\beta_0(a_0)=0$, $y_1(x)={\mathcal O}(1-x^2)$ 
when $x\to\pm 1$. These facts mean that $y_1(x)=xp_1(x^2)(1-x^2)$, where $p_1(x)$ is a polynomial 
of degree $q-1$. From these arguments and using induction over $n$ in formula (\ref{yn}) 
we can easily obtain thesis (iii).

\section{The first few terms of the expansion}
\label{sec-ex}
For $\epsilon=0$, Eq. (\ref{lienard}) has a continuum of periodic solutions: $u(t)=ay_0(t/a)$ 
for any value of $a$. Hence, $y_0(1)=0$ does not impose any restriction over the parameter $a$. 
we determine $a_0,...,a_{N-1}$ and $y_1(x)$,...,$y_N(x)$ by means of the above introduced algorithm.
First, we take $y_0(x)=\sqrt{1-x^2}$. Then, for $n=0,1,2,...,N-1$, solve the equation
$\beta_n(a_0,a_1,...,a_n)=0$ for $a_n$ and then compute $y_{n+1}(x)$. We give
below the details of this algorithm for $N=5$. 

\noindent                                                        
{\bf Definition 4.}
For convenience we define, for $n=2,4,6,....,$ the integrals
$$
\gamma_n(a_0,a_2,...,a_{n-2})\equiv\int_{-1}^1 f(a_0 x)y_n(x)dx.
$$

\noindent
\underline{\bf The order $n=0$}. We calculate $a_0$ and $y_1(x)$ as follows.
Compute the positive $a_0$ solutions of
$$
\beta_0(a_0)=\int_{-1}^1f(a_0x)\sqrt{1-x^2}dx=0.
$$
From (\ref{yn}) we obtain the expression (\ref{yuno}) for $y_1(x)$:
$$
y_1(x)=-{1\over y_0(x)}\int_{-1}^x f(a_0 t)y_0(t)dt.
$$

\noindent
\underline{\bf The order $n=1$}. We know that $a_1=0$ and we calculate 
$y_2(x)$ from (\ref{yn}) with $n=2$:
$$
y_2(x)=-{1\over y_0(x)}\left[{1\over 2}y_1^2(x)+\int_{-1}^x f(a_0 t)y_1(t)dt\right].
$$

\noindent
\underline{\bf The order $n=2$}.  We calculate $a_2$ and $y_3(x)$. 
From (\ref{betan}) with $n=2$ we have
$$
\beta_2(a_0,0,a_2)=\int_{-1}^1 f(a_0 x)y_2(x)dx+
a_2\int_{-1}^1 xf'(a_0 x)y_0(x)dx=\gamma_2(a_0)+a_2\beta_0'(a_0)=0
$$
Then, 
\begin{equation}
\label{ados}
a_2=-{\gamma_2(a_0)\over\beta_0'(a_0)}.
\end{equation}
From (\ref{yn}) with $n=3$:
$$
y_3(x)=-{1\over y_0(x)}\left[y_1(x)y_2(x)+\int_{-1}^x f(a_0 t)y_2(t)dt+a_2 
\int_{-1}^x tf'(a_0 t)y_0(t)dt\right].
$$

\noindent
\underline{\bf The order $n=3$}. We know that $a_3=0$ and we calculate 
$y_4(x)$ from (\ref{yn}) with $n=4$:
$$
y_4(x)=-{1\over y_0(x)}\left[y_1(x)y_3(x)+{1\over 2}y_2^2(x)+
\int_{-1}^x f(a_0 t)y_3(t)dt+a_2\int_{-1}^x tf'(a_0 t)y_1(t)dt\right].
$$

\noindent
\underline{\bf The order $n=4$}.  We calculate $a_4$ and $y_5(x)$. From (\ref{betan}) with $n=4$ we have
$$
\beta_4(a_0,0,a_2,0,a_4)=\gamma_4(a_0,a_2)+a_2\gamma_2'(a_0)+{a_2^2\over 2}\beta_0''(a_0)+a_4\beta_0'(a_0)=0
$$
Then, 
\begin{equation}
\label{acuatro}
a_4=-{\gamma_4(a_0,a_2)+a_2\gamma_2'(a_0)+{a_2^2\over 2}\beta_0''(a_0)\over\beta_0'(a_0)}.
\end{equation}
From (\ref{yn}) with $n=5$:
\begin{eqnarray}
y_5(x) & = & -{1\over y_0(x)}\left[y_1(x)y_4(x)+y_2(x)y_3(x)+\int_{-1}^x f(a_0 t)y_4(t)dt+a_2 
\int_{-1}^x tf'(a_0 t)y_2(t)dt+ \right.  \nonumber \\ 
&& \left. 
a_4\int_{-1}^x tf'(a_0 t)y_0(t)dt+{a_2^2\over 2} \int_{-1}^x t^2f''(a_0 t)y_0(t)dt\right].
\end{eqnarray}

\noindent
{\bf Remark 1.}
The integral $\beta_0(a_0)$ coincides with the
first Melnikov function or, equivalently, with the Abelian integral \cite{arnold}
defined for the perturbed Hamiltonian system (\ref{liena}) whose level curves 
at $\epsilon=0$ are given by the set of circles $x^2+y^2=a^2$. Thus,
the equation $\beta_0(a_0)=0$ is a nonlinear equation for $a_0$ with none, 
one or several solutions, whereas the remaining equations $\beta_n(a_0,a_1,...,a_n)=0$ 
with $n=1,2,..,N-1$ are linear equations in $a_n$ with a unique solution for every $a_0$ solution of
$\beta_0(a_0)=0$. Then, the number of positive solutions $a_0$ of $\beta_0(a_0)=0$ determine the 
maximal number of limit cycles emerging from the period annulus when it is perturbed
and the ${\mathcal O}(1)$ approximation to their amplitude. 
The remaining equations $\beta_n(a_0,a_1,...,a_n)=0$ with $n=1,2,..,N-1$ determine recursively 
the coefficients $a_1$,...,$a_{N-1}$, which are the pertubative correction to the first order amplitudes $a_0$. 
In this way, for every solution of $\beta_0(a_0)=0$, we completely determine the amplitude $a$ 
and the form $y(x)$ of the limit cycle solutions of (\ref{liena}) up to the order ${\mathcal O}(\epsilon^N)$: 
$$
a\simeq a^{(N)}\equiv \sum_{n=0}^{N-1} a_n\epsilon^n,\hskip 15mm
y(x)\simeq y^{(N)}(x)\equiv \sum_{n=0}^{N} \epsilon^ny_n(x).
$$
If $f(x)$ is a polynomial of degree $q$ in $x^2$, then $\beta_0(a_0)$ is a 
polinomial in $a_0^2$ of degree $q$. Then, for small $\epsilon$, the maximum possible number 
of limit cycles of (\ref{liena}) is $q$, as it has been conjectured by 
Lins, Melo and Pugh (LMP-conjecture) \cite{lins}.

\noindent
{\bf Remark 2.}
The approximate solution $y^{(N)}(x)$ with amplitude $a^{(N)}$  
has the same symmetries than the exact solutions of Eq. (\ref{liena}): 
the {\it inversion symmetry} $(x,y)\leftrightarrow (-x,-y)$ and the
{\it parameter inversion symmetry} $(\epsilon,x,y)\leftrightarrow (-\epsilon,-x,y)$. 
These symmetries are not destroyed at the perturbative level.

\noindent
{\bf Remark 3.}
From thesis (iii) of Theorem 1, 
when $f(x)$ is a polynomial of degree $q$ in $x^2$ we have that
$$
y^{(N)}(x)=\sqrt{1-x^2}+x(1-x^2)\sum_{n=0}^{\lfloor N/2\rfloor}
p_{2n+1}(x^2)\epsilon^{2n+1}+(1-x^2)^{3/2}\sum_{n=1}^{\lfloor N/2\rfloor} p_{2n}(x^2)\epsilon^{2n},
$$
where $p_n(x)$ is a polynomial of degree $nq-1$.

\section{Examples}

We perform the calculations proposed in Section 4 for two concrete examples,
which are particular cases of the families $1$ and $3$ worked out in \cite{lopezruiz1}.

\noindent
{\bf Example 1.} The van der Pol oscillator is given for $f(x)=x^2-1$.
This system has a unique limit cycle, which is stable for $\epsilon>0$.
Hence, the only root of $\beta_0(a)$ is $a_0=2$. For this value of $a_0$, 
an approximate form of the limit cycle for small $\epsilon$ is
$y^{(5)}(x)=\sum_{n=0}^5\epsilon^n y_n(x)$ with
$$
y_0(x)=\sqrt{1-x^2}, \hskip 2cm y_1(x)=x(1-x^2),
$$
$$
y_2(x)={1\over 12}(1+2x^2)(1-x^2)^{3/2}, \hskip 1cm y_3(x)={1\over 72}x^3(6x^2-5)(1-x^2),
$$
$$
y_4(x)={1\over 4320}(-4-23x^2+213x^4-156x^6)(1-x^2)^{3/2}
$$
and
$$
y_5(x)={1\over 1036800}x^3(10385-43794x^2+41424x^4-10560x^6)(1-x^2).
$$

An approximate form of its amplitude for small $\epsilon$ is $a(\epsilon)=a^{(6)}+{\mathcal O}(\epsilon^8)$, with
\begin{equation}
\label{aa}
a^{(6)}=2+{1\over 96}\epsilon^2-{1033\over 552960}\epsilon^4+{1019689\over 55738368000}\epsilon^6.
\end{equation}

We integrate Eq. (\ref{eq3}) by a Runge-Kutta method in order
to obtain the limit cycle. This curve is plotted in a continuous trace in Fig. 1(a-b)
for $\epsilon=0.8$ and $\epsilon=2$, respectively. The approximated limit
cycle $u^{(5)}(x)=a^{(5)}y^{(5)}(x/a^{(5)})$ is also plotted in those figures with a discontinuous trace
for the same values of $\epsilon$.
Let us remark that, in this case, even up to $\epsilon=3$,
the approximation $u^{(5)}(x)$ to the limit cycle is very good.
Our calculation (\ref{aa}) agrees with the computational calculation 
of the 'exact' amplitudes given in Table 1,
and also with the calculations on this system presented in \cite{giacomini1}.

\noindent
{\bf Example 2.} The same process is performed for $f(x)=5x^4-9x^2+1$.
In this case, the system has two limit cycles, one stable and the other unstable.
The polynomial $\beta_0(a)$ has two positive roots:
$a_0=\sqrt{9-\sqrt{41}\over 5}=0.720677$ (unstable limit cycle for $\epsilon>0$)
and $\bar a_0=\sqrt{9+\sqrt{41}\over 5}=1.75517$ (stable limit cycle for $\epsilon>0$).

For the first limit cycle we have $y^{(3)}(x)=y_0(x)+\epsilon y_1(x)+\epsilon^2 y_2(x)+\epsilon^3 y_3(x)$ with
$$
y_0(x)=\sqrt{1-x^2},
$$
$$
y_1(x)=x(x^2-1)\left(1+{9\sqrt{41}-61\over 15}x^2\right),
$$
$$
y_2(x)=(1-x^2)^{3/2}(0.06586+0.13173x^2-0.05241x^4+0.00505x^6)
$$
and
$$
y_3(x)=x^3(1-x^2)(0.06081-0.11246x^2+0.05384x^4-0.00999x^6+0.00006x^8).
$$

An approximate expression of its amplitude for small $\epsilon$ is
$$
a(\epsilon)=0.720677+0.00390888\;\epsilon^2+{\mathcal O}(\epsilon^4).
$$

For the second limit cycle we have $y^{(3)}(x)=y_0(x)+\epsilon y_1(x)+
\epsilon^2 y_2(x)+\epsilon^3 y_3(x)$ with
$$
y_0(x)=\sqrt{1-x^2},
$$
$$
y_1(x)=x(x^2-1)\left(1-{9\sqrt{41}+61\over 15}x^2\right),
$$
$$
y_2(x)=(1-x^2)^{3/2}(0.98791+1.97583x^2-2.71374x^4+6.2545x^6)
$$
and
$$
y_3(x)=x^3(1-x^2)(0.846907-2.30045x^2-1.11829x^4-25.2371x^6+28.2651x^8).
$$

An approximate form of its amplitude for small $\epsilon$ is
$$
a(\epsilon)=1.75517-0.0178803\;\epsilon^2+{\mathcal O}(\epsilon^4).
$$

The comparison between the 'exact' and the approximated limit cycles
can be seen in the plots of Fig. 2(a-b) for $\epsilon=0.8$ and $\epsilon=2$
and Fig. 3(a-b) for $\epsilon=0.4$ and $\epsilon=0.8$.

\section{Conclusions}

Periodic self-oscillations can arise in nonlinear systems. 
These are represented by isolated closed curves in phase 
space that we call limit cycles.
The knowledge of the number, amplitude and shape of these solutions
in a general nonlinear equation is an unsolved problem.

In this work, 
a recursive algorithm to approximate the form $y(x)$ and the amplitude $a$
of limit cycles in the weakly nonlinear regime ($\epsilon\rightarrow 0$)
of Li\'enard equations,
$\dot x=y$, $\dot{y}+\epsilon f(x)y+x=0$, with $f(x)$ an even function, 
has been presented. The symmetries of the exact limit cycles are maintained 
at the perturbative level.
Several examples showing the application of this scheme 
have been given, and the accuracy of the method in comparison with the direct
numerical integration has also been tested. 
In particular, we have explicitly obtained the ${\mathcal O}(\epsilon^8)$ 
approximation for the amplitude of the van der Pol limit cycle.

{\bf Acknowledgements:}  R. L-R. acknowledges some financial support 
from the spanish research project FIS2004-05073-C04-01. 
J. L. L. acknowledges the financial support of the {\it Direcci\'on General de Ciencia y
Tecnologia (REF. MTM2004-05221)}.

%\newpage

%\newpage
\vskip 2 cm
\noindent{\bf Table 1.} The value $a_{T}$ represents the approximated
amplitude $a(\epsilon)$ of the van der Pol limit cycle
obtained from (\ref{aa}) for the indicated values of $\epsilon$.
The value $a_{E}$ represents the amplitude $a$ obtained by integrating directly
the system with a Runge-Kutta method.

\begin{center}
\begin{normalsize}
\begin{flushleft}
\begin{tabular}{|p{.6cm}|p{2cm}|p{2cm}|p{2cm}|p{2cm}|l|l|}
\hline $\epsilon$ & $0.1$ & $0.2$ & $0.3$ & $0.4$ & $0.5$ \\
\hline $a_{T}$&2.000104&2.000414&2.000922&2.001619&2.002488\\
\hline $a_{E}$&2.00010&2.00041&2.00092&2.00161&2.00248\\
\hline
\hline $\epsilon$ & $0.6$ & $0.7$ & $0.8$ & $0.9$ & $1$\\
\hline $a_{T}$&2.003509&2.004658&2.005906&2.007222&2.008569\\
\hline $a_{E}$&2.00351&2.00466&2.00591&2.00724&2.00862\\
\hline
\end{tabular}
\end{flushleft}
\end{normalsize}
\end{center}

\vskip 2 cm
\begin{center} {\bf Figures} \end{center}

{\bf Fig. 1a-b}:
Exact limit cycle (continuous line) and approximated limit cycle (discontinuous line)
up to order ${\mathcal O}(\epsilon^5)$ for the van der Pol system with
(a) $\epsilon=0.8$ and (b) $\epsilon=2$.

{\bf Fig. 2a-b}:
Exact limit cycle (continuous line) and
approximated limit cycle (discontinuous line) up to order
${\mathcal O}(\epsilon^3)$ for the unstable limit cycle
of the example 2 with (a) $\epsilon=-0.8$ and (b) $\epsilon=-2$.

{\bf Fig. 3a-b}:
Exact limit cycle (continuous line) and
approximated limit cycle (discontinuous line) up to order
${\mathcal O}(\epsilon^3)$ for the stable limit cycle
of the example 2 with (a) $\epsilon=0.4$ and (b) $\epsilon=0.8$.

\newpage

\begin{figure}[]
\includegraphics[angle=0, width=15cm]{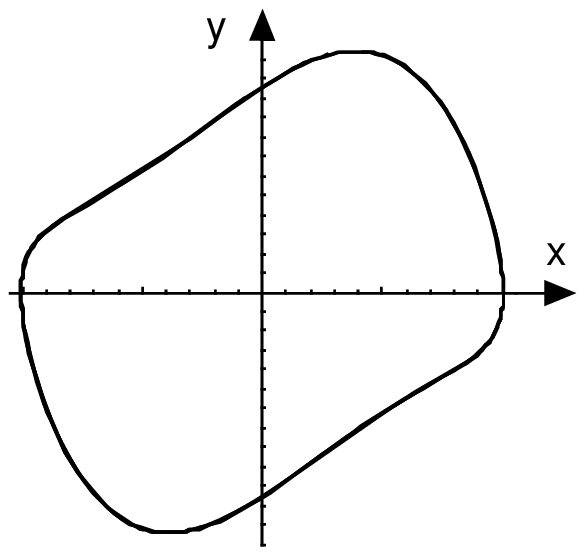}
\label{fig1a}
\end{figure}

\begin{figure}[]
\includegraphics[angle=0, width=15cm]{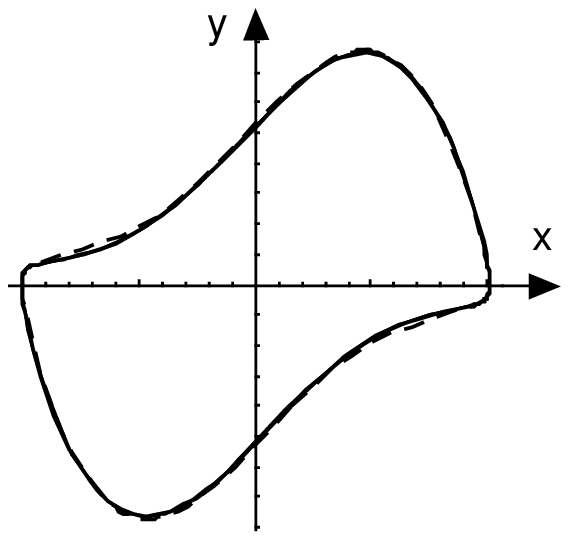}
\label{fig1b}
\end{figure}

\newpage
\begin{figure}[]
\includegraphics[angle=0, width=15cm]{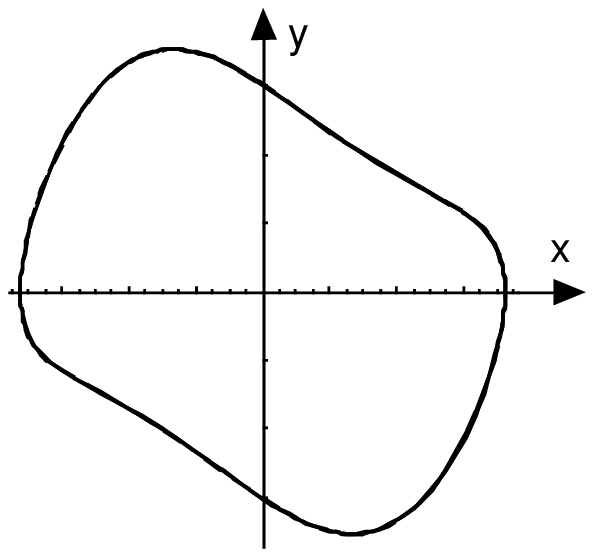}
\label{fig2a}
\end{figure}

\begin{figure}[]
\includegraphics[angle=0, width=15cm]{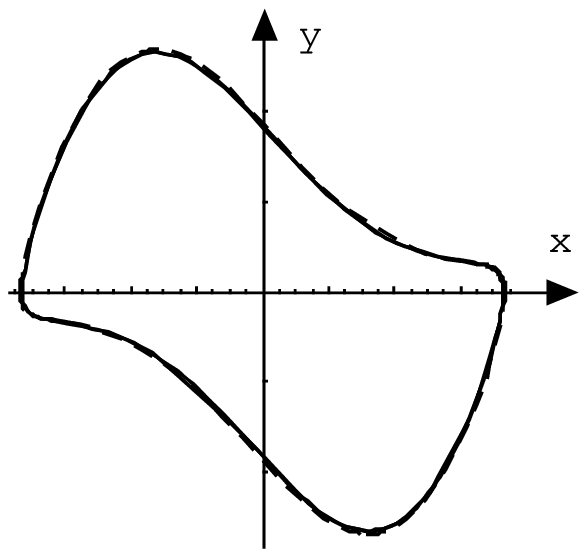}
\label{fig2b}
\end{figure}

\begin{figure}[]
\includegraphics[angle=0, width=15cm]{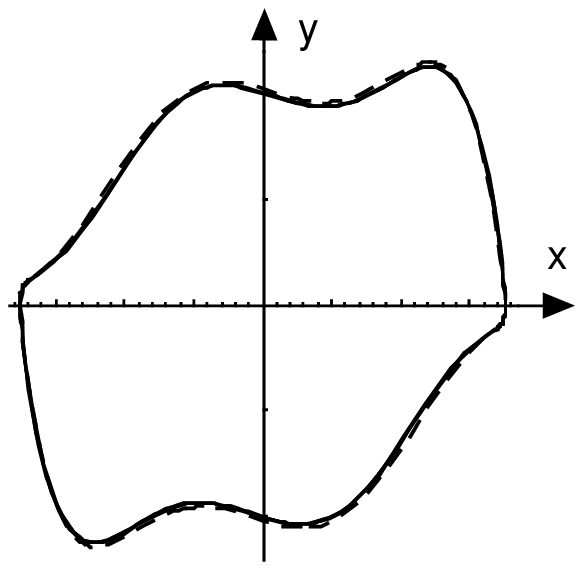}
\label{fig3a}
\end{figure}

\begin{figure}[]
\includegraphics[angle=0, width=15cm]{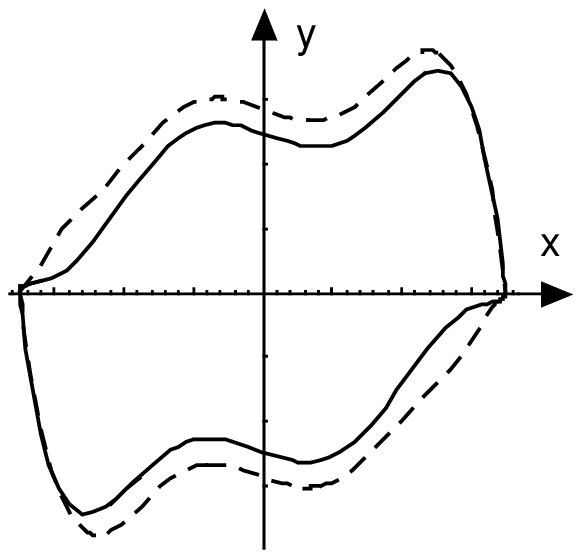}
\label{fig3b}
\end{figure}

\end{document}